\documentclass{emulateapj}

\usepackage{epsf}
\usepackage{graphics}
\usepackage{color}

\newcommand{\be}{\begin{equation}}
\newcommand{\ee}{\end{equation}}

\newcommand{\ax}{$\alpha_{\rm X}$}

\newcommand{\aox}{$\alpha_{\rm ox}$}

\newcommand{\rb}[1]{\raisebox{1.5ex}[-1.5ex]{#1}}
\newcommand{\msun}{$M_{\odot}$}

\newcommand{\plm}{$\pm$}

\newcommand{\swift}{{\it Swift}}
\newcommand{\xmm}{{\it XMM-Newton}}
\newcommand{\chandra}{{\it Chandra}}
\newcommand{\wpvs}{{WPVS~007}}

\shorttitle{Swift detection of WPVS 007}
\shortauthors{Grupe et al.}

\begin{document}


\def\etal{{\it et\thinspace al.}\ }
\def\alp{{$\alpha$}\ }
\def\al2{{$\alpha^2$}\ }


\title{First detection of hard X-ray photons in the soft X-ray 
transient Narrow-Line Seyfert 1 galaxy
WPVS 007: The X-ray photon distribution observed by \swift
}


\author{Dirk Grupe\altaffilmark{1},
Karen M. Leighly\altaffilmark{2},
Stefanie Komossa\altaffilmark{3},
}

\altaffiltext{1}{Department of Astronomy and Astrophysics, Pennsylvania State
University, 525 Davey Lab, University Park, PA 16802; email: grupe@astro.psu.edu}

\altaffiltext{2}{Homer L. Dodge
Department of Physics and Astronomy, University of Oklahoma, 
440 West Brooks Street, Norman, OK 73019; email: leighly@nhn.ou.edu}

\altaffiltext{3}{Max-Planck-Institut f\"ur extraterrestrische Physik, Giessenbachstr., D-85748 Garching,
Germany; email: skomossa@mpe.mpg.de}




\begin{abstract}
We report on the first detection of hard X-ray photons ($E>$2.5 keV)  in
the X-ray transient Narrow-Line Seyfert 1 galaxy WPVS 007 
which was the AGN with the softest X-ray spectrum during the 
ROSAT All-Sky Survey. 
The AGN is clearly detected at a level of about 2$\times 10^{-17}$ W m$^{-2}$
in the observed 0.3-10.0 keV band
by \swift\ in a 50 ks observation in 2007 September. For the first time
since the ROSAT All-Sky Survey observation in 1990 it was possible 
to derive an
X-ray photon distribution by adding together all \swift\ observations
 that have been performed so far 
(85.5 ks in total). This photon distribution is consistent with an
X-ray spectrum  of an AGN with a partial covering absorber with a column
density in the order of $\sim 1\times10^{23}$ cm$^{-2}$ and a covering 
fraction of about 90\%.  A
comparison with the 2002 \chandra\ data suggests that WPVS 007 has become
brighter by a factor of about 4. The \swift\ data also 
suggest
that the absorber which is causing the current low-state may have started to
disappear. This disappearance is indicated by a significant change in the hardness
ratio from a very soft X-ray state during the 2005 October to 2007 January observations
to a rather hard X-ray state in the 2007 September observations.
 In the UV, WPVS 007 seems to become fainter by up to 0.5 mag over
the last two years.
The optical to X-ray spectral slope derived from the spectral
energy distribution is \aox=2.5 which classifies WPVS 007 as an X-ray weak
AGN. After correcting for reddening and X-ray absorption, 
\aox\ becomes 1.9 and the luminosity in the Big-Blue-Bump is log
$L_{\rm BBB}$=37.7 [W], which translates into an Eddington ratio
 $L/L_{\rm Edd}\approx 1$.
\end{abstract}


\section{Introduction}

The Narrow-Line Seyfert 1 galaxy (NLS1) \wpvs\ 
\citep[1RXS J003916.6$-511701$, RBS 0088;
$\alpha_{2000}$ = $00^{\rm h} 39^{\rm m} 15.^{\rm s}8$, 
$\delta_{2000}$ = $-51^{\circ} 17' 03 \farcs 0$, z=0.029; ][]{wamsteker85,grupe95}
is a unique X-ray transient AGN. 
During the ROSAT All-Sky Survey \citep[RASS, ][]{voges99},
   when WPVS 007 was in its X-ray high-state, 
   the X-ray-to-optical flux ratio was log 
   $f_{\rm x}/f_{\rm o}$\footnote{$f_{\rm x}/f_{\rm o}$ is defined by
   \citet{beu99} as log($f_{\rm x}/f_{\rm o}$)= log CR + 0.4 V - 5.86, with CR =
the ROSAT PSPC count rate in the 0.1-2.0 keV range.}=0.22
   which is typical
   for a Seyfert 1 galaxy \citep[e.g. ][]{beu99, maccacaro88}. 
   Nevertheless, in all X-ray follow-up observations between 1993 and 2007 
   using ROSAT, \chandra\, and \swift\ WPVS 007 has been found in an extreme  
 X-ray low-state that made 
   WPVS 007 almost vanish from the X-ray sky \citep{gru01, grupe07}. It had
   been last detected in 2002 August by Chandra with $1\times 10^{-3}$ 
  ACIS-S counts s$^{-1}$ \citep{vaughan04}.
 While this strange transient behavior had been a mystery for a 
 decade,   
\citet{leighly05} and \citet{leighly08}  discovered that
it is probably absorption that causes the extreme low-state in this AGN.
They found in a 2003 November {\it FUSE} observation that the UV
spectrum of WPVS 007 had changed dramatically compared with HST
observations obtained in 1996 July    \citep{goodrich00,
constantin03}:   the {\it FUSE} observation revealed the emergence of
a  BAL flow.   \citet{leighly08} concluded that WPVS 007 is a
low-luminosity, low black hole   mass cousin of Broad Absorption Line
Quasars (BALQSOs). The relatively short timescale for the development
of the BAL inferred in WPVS 007 suggests that it could re-brighten at
any time on relatively short timescales. 
We started a monitoring campaign using \swift\ in
2005 October \citep{grupe07}. The purpose of this campaign is to detect a
possible rebrightening and therefore the disappearance of the absorber. By the end
of the second year of monitoring WPVS 007 with \swift\ in 2007 January, only a
3$\sigma$ upper limit of the observed 
X-ray flux could be derived at a level of $2.3\times
10^{-17}$ W m$^{-2}$ in the rest-frame 0.2-2.0 keV band. 

\citet{brandt00} found that about 10\% of optically selected quasars appear
to be X-ray weak with an optical to X-ray spectral slope 
\aox\footnote{The X-ray loudness is defined by \citet{tananbaum79} as
\aox=--0.384 log($f_{\rm 2keV}/f_{2500\AA}$).}$>$2.0. X-ray weakness can either
be intrinsic, such as seen in e.g. PHL 1811 
\citep[][]{leighly07}, or caused by intrinsic absorption.
Low X-ray flux states 
have been found in several NLS1s, such as Mrk 1239, 1H 0707--495,
and most recently in Mrk 335 \citep{grupe04b, boller02, gallo04, 
grupe07b}. In Mrk 335 \citet{grupe07b} showed that this dramatic drop in the
X-ray flux goes along with a strong change in the X-ray spectrum. One of the
possible scenarios to explain the X-ray spectra of low-state NLS1s is a partial
covering absorber. An alternative solution to the strange looking low-state
spectra of these NLS1s is reflection as suggested for example by
\citet{fabian89, fabian04}. The X-ray spectra by themselves do not
provide a definite 
solution to distinguish between both models because with the quality 
of the spectra typically obtained and the number of free model parameters,
both models fit the X-ray spectra 
well and they are statistically indistinguishable. 
In addition, the cases of several well-studied AGN
have shown that low X-ray flux states are temporary
events and not necessarily persistent \citep[e.g.][]{komossa97, 
guainazzi98, costantini00, komossa01, bianchi05,
grupe07b, grupe08, ballo08a}.


The \swift\ Gamma-Ray Burst  explorer mission \citep{gehrels04} was
launched on 2004 November 20.  With  its  simultaneous
multi-wavelength capacity and its scheduling flexibility, \swift\ has
shown itself to be the ideal observatory to study highly variable
objects like AGN.  Even though \swift's main purpose 
is to detect and observe Gamma-Ray bursts, \swift\ has also done an
excellent job in observing AGN as fill-in targets or
Target-of-Opportunity (ToO) observations \citep[e.g. ][]{markwardt05,
  grupe06, grupe07, grupe07b, kataoka07, sambruna07, giommi07}. 
\swift's UV/Optical Telescope  \citep[UVOT, ][]{roming04} and X-ray
Telescope \citep[XRT, ][]{burrows04} cover the electromagnetic
spectrum between 6500\AA\ on the low energy side  to 10 keV at the
high energy end simultaneously.  
 
We started the WPVS 007 monitoring campaign again in 2007 July. On 2007 August 10
a failure of one of the \swift\ gyros forced the spacecraft into a safe-hold
\citep{gehrels07}. During the spacecraft recovery process, \swift\ was
slewing only around a small circle in the sky that included the
position of WPVS 007.  We took advantage of this limited slewing
capacity and obtained about 50ks of X-ray observations of the field of
WPVS 007 in 2007 September.  During these observations we were not
only able to clearly detect WPVS 007 in X-rays, but even more
importantly, we discovered a significant number of photons at hard
energies above 2 keV, and derived a photon distribution that gives us
information for the first time since the RASS about the X-ray
spectrum.  In this paper we present and discuss these new
observations. 

The outline of this paper is as follows: in \S\,\ref{observe} we describe the
\swift\  observations and the data reduction, in \S\,\ref{results} we 
present the results of the \swift\ XRT data,
and in \S\,\ref{discuss} we discuss the results. 
Throughout the paper spectral indexes are denoted as energy spectral indexes
with
$F_{\nu} \propto \nu^{-\alpha}$. Luminosities are calculated assuming a $\Lambda$CDM
cosmology with $\Omega_{\rm M}$=0.27, $\Omega_{\Lambda}$=0.73 and a Hubble
constant of $H_0$=75 km s$^{-1}$ Mpc$^{-1}$ using a luminosity distances D=118 Mpc
given by \citet{hogg99}. All errors are 1$\sigma$ unless stated otherwise.

\section{\label{observe} Observations and data reduction}

Table\,\ref{xrt_log} presents the new
\swift~XRT observations of \wpvs\ starting on 2007
July 15, including the start and end times and the total exposure
times. A list of the previous  \swift\ observations between 2005
October and 2007 January is given in \citet{grupe07}.  
Due to the spacecraft recovery UVOT data
were only taken before the safe-hold on 2007 August 10, and after the
turn on of the UVOT by the end of 2007 October.
The XRT was operating in photon counting mode \citep{hill04} and the
data were reduced by the task {\it xrtpipeline} version 0.10.4., 
which is included in the HEASOFT package 6.1. 
For creating an image of the 2007 September observations we used {\it XSELECT}
version 2.4. We ran the source detection algorithm {\it detect} in {\it XIMAGE}
version 4.4 with a signal-to-noise level set to 3.

In order to obtain a source photon distribution from the WPVS 007
data, we co-added the data from all of the \swift\ observations 
(85.5 ks in total) of WPVS
007  \citep[except for segment005
for which the background was too high; ][]{grupe07}. From
these co-added data source photons were extracted   
using {\it XSELECT}  from a circle with r=$23\farcs4$ and background
photons from a source-free region close by with a radius r=$95\farcs0$.
Thirty six net source photons were obtained, too few to fit using the
usual $\chi^2$ statistics.  Because the source photons account for only
70\% of the total, even Cash statistics cannot be applied  
in {\it XSPEC} \citep{arnaud96} to fit the photon distribution. 
Therefore, the spectrum has to be reconstructed by hardness ratio
analysis. The hardness ration is defined as HR=(H-S)/(H+S) with S and 
H are the number of source photons in the 0.3-2.0 and 2.0-10.0 keV
energy bands, respectively. The hardness ratios were calculated using
the program BEHR as described in \citet{park06} which uses Bayesian
statistics which is required for the small number of photons in WPVS
007. The absorption column density of the z=0 absorber was set to the
Galactic value  \citep[$N_{\rm H}=2.84\times10^{20}$ cm$^{-2}$ ][]{dic90}.

The UVOT data  were reduced and analyzed as described in \citet{grupe07}. 
Due to UVOT calibration and software changes in June 2007 \citep{poole07}
we reanalyzed the UVOT data previously obtained by \swift\ with a
source extraction radius of $5\farcs0$. Primarily the values in the UV
filters changed in the order of about 0.3 mag. For consistency we 
list the results and exposure times in Table\,\ref{uvot_log}
of all \swift\ UVOT observation, including those prior to June 2007,
previously published in \citet{grupe07}.

\section{\label{results} Results}

\subsection{X-ray detection and analysis}

\begin{figure}
\epsscale{0.6}
\plotone{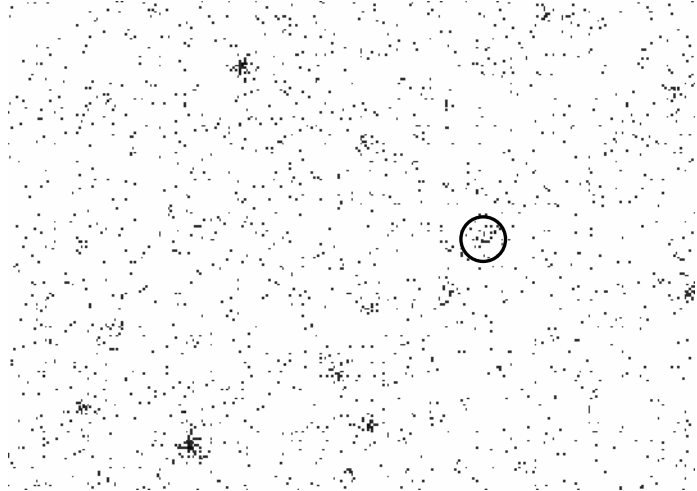}
\caption{\label{wpvs007_image} Image of the 50 ks observation in 2007 September
of the field of WPVS 007 in the 0.3-10.0 keV energy band.
 The black circle marks the position of WPVS 007.
}
\end{figure}

\begin{figure}
\epsscale{1.0}
\plottwo{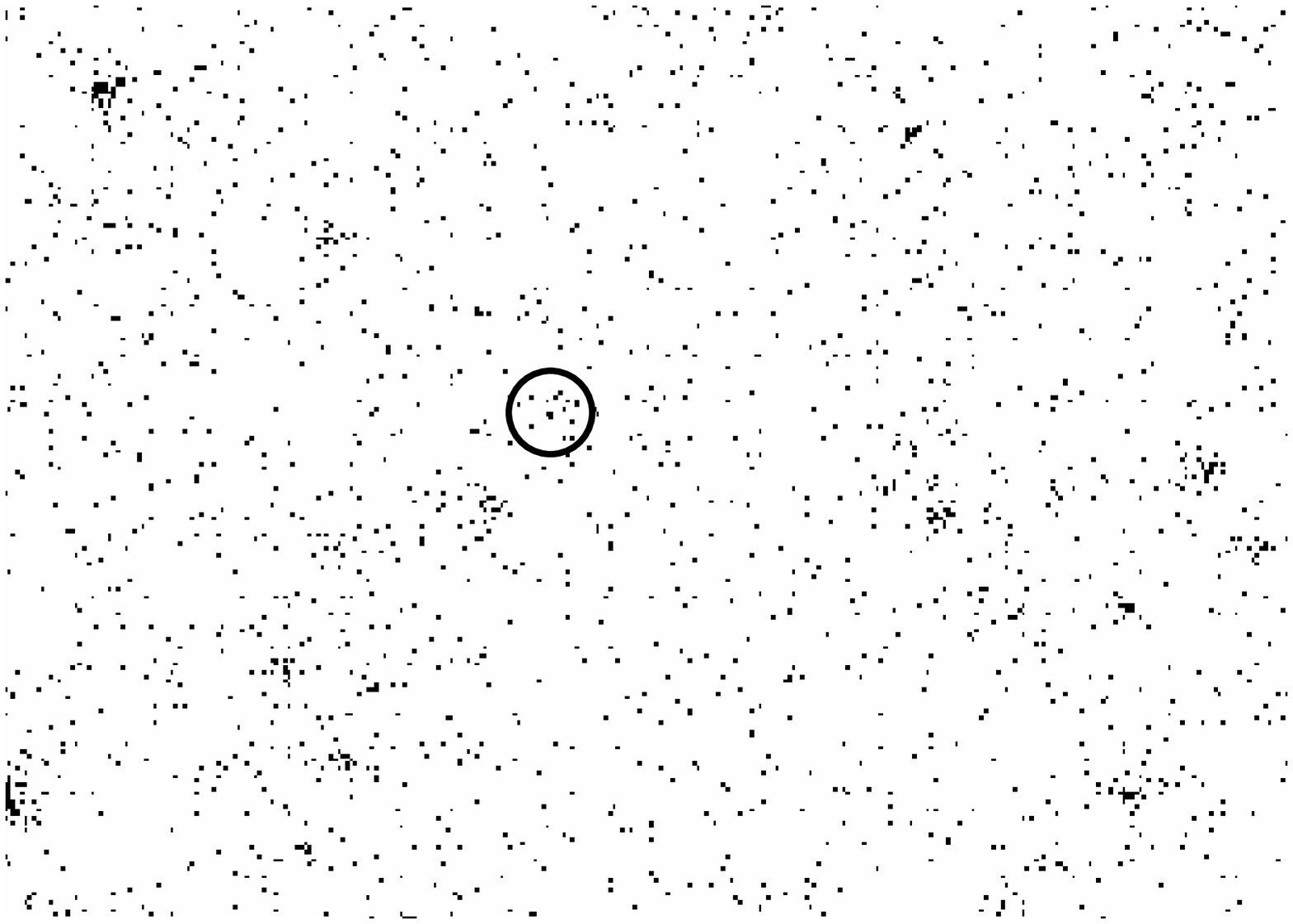}{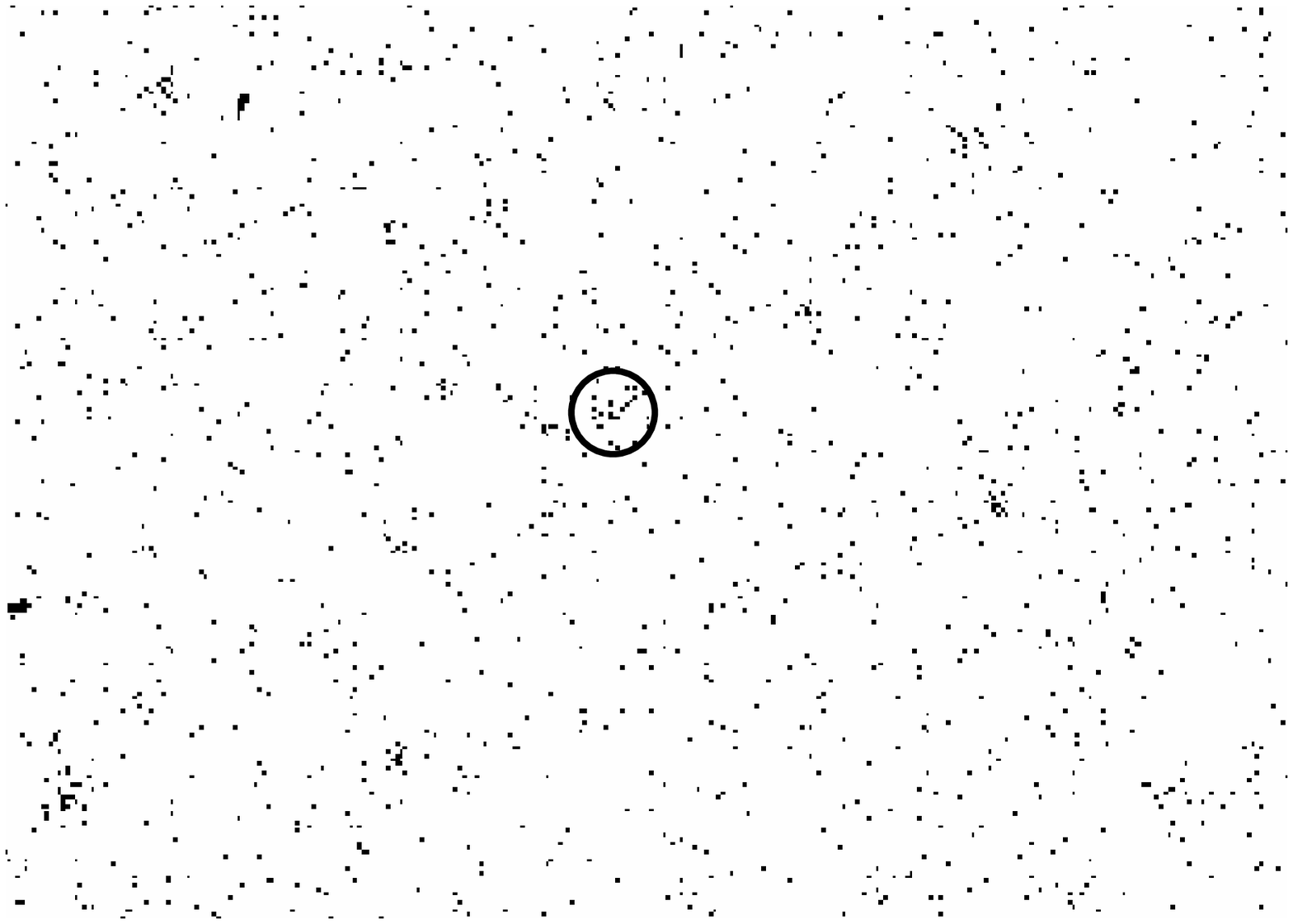}
\caption{\label{wpvs007_image_all} Images of the 0.3-2.0 keV and 2.0-10.0 keV
energy ranges (left and right panels, respectively) from all Swift observations
obtained on the field of WPVS 007 so far, 85.5 ks in total. 
 The black circle marks the position of WPVS 007.
}
\end{figure}

Figure\,\ref{wpvs007_image} displays the image of the field of \wpvs\ during the
50ks observations of 2007 September. The circle marks the position of \wpvs. 
Using the {\tt detect} algorithm in  {\it XIMAGE}, the
AGN is clearly detected at a level of (5.1\plm1.2)$\times 10^{-4}$ 
XRT counts s$^{-1}$.  Adding all XRT data together that have
been obtained since 2005 October (85.5 ks in total) results in a 5$\sigma$
detection with (4.5\plm0.9)$\times 10^{-4}$ counts s$^{-1}$. 
Figure\,\ref{wpvs007_image_all} displays the field of WPVS 007 in the 
soft 0.3-2.0
keV and hard 2.0-10.0 keV bands on the left and right panels, respectively. While
there is only a marginal detection at the 2$\sigma$ level in the soft X-ray
band, the AGN is clearly detected at the 4.5$\sigma$ level in the hard X-ray
band.  Note, that in the 2005 October to 2007 January data there is an enhanced
number of counts at the position of WPVS 007 (Table\,\ref{xrt_stat}).
However, there is no obvious clustering of photons at that position. Therefore we
only report a 3$\sigma$ upper limit for this time period.  

The biggest surprise of the \swift\ observations is the 
detection of \wpvs\ at
hard X-ray energies. From the whole 85.5 ks data set
we detected 24 source photons with energies $E>$2.5 keV out of a total of 36
source photons. 
 This is surprising because during the RASS WPVS 007 was {\bf the} AGN with the
steepest X-ray spectrum (\ax=8.7) and with no X-ray photon with energies
above 0.5 keV \citep{grupe95}.  During the 10ks \chandra\ observation on 
2002 August 02 out of
10 photons only one was detected above 2 keV, at 2.3 keV.

Table\,\ref{xrt_stat} lists the number counts, count rates, and
hardness ratios
for the 2005 October to 2007 January, 2007 September, and 2005
October to 2008 January observations.  The Hardness ratios between the
2005 October to 2007 January observations with the 2007 September indicated a
dramatic change in the X-ray spectrum. While the hardness ratio in the 2005 October
to 2007 January observations is soft with HR=$-0.49^{+0.23}_{-0.50}$ which is
consistent with the hardness ratio during the 2002 Chandra observation
(HR=$-0.86^{+0.12}_{-0.14}$), WPVS 007 had become significantly harder with
HR=+0.63\plm0.20 in the 2007 September observations. Although comparing hardness
ratios of different detectors is difficult, X-ray CCDs are similar enough to
tell if the majority of photons is in the hard or the soft bands. 

\begin{figure}
\epsscale{0.4}
\plotone{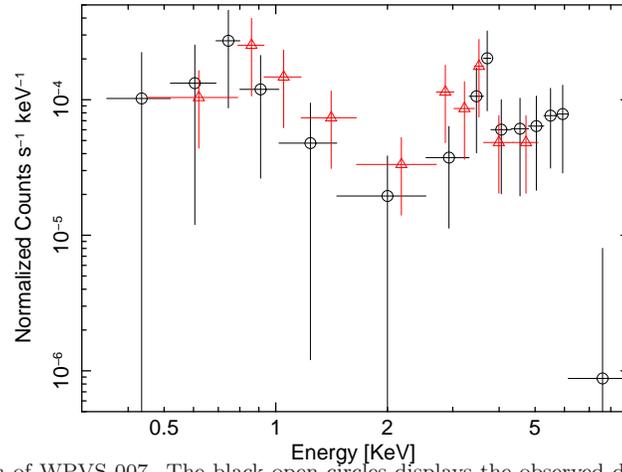}
\caption{\label{wpvs007_xspec} X-ray photon distribution 
of WPVS 007. The black open circles displays the observed data, 
containing all \swift\ observations with a total exposure time of 85.5 ks.
The red open triangles 
 display a simulated 85 ks spectrum using a partial covering
absorber with a single power law 
model as described in the text to represent the observed photon distribution. 
For clarity reasons the spectra were binned with 3 photons per bin.
}
\end{figure}

As listed in Table\,\ref{xrt_stat},
the total number of source photons in the combined 85 ks \swift\
exposure\footnote{Note, as mentioned in \citet{grupe07}
 the segment 005 data were excluded from the analysis due to a high detector
 background during the observation.} is only about 36. 
Although the number of source photons is small, 
this photon distribution can be used to examine
 what spectral model best represents the
 current low-state X-ray spectrum of \wpvs\/. 
 Figure\,\ref{wpvs007_xspec} shows the observed  and simulated photon
 distribution (black circles and red triangles, respectively).
 For presentation purposes only the spectrum was rebinned with 3 
 photons per bin.
 Because only 68\% of the total number of photons in the
 non-background subtracted spectrum are source photons, no spectral
 fitting of the data is possible. In order to determine what spectral model can
 represent the observed photon distribution we measured the number of source
 photons in different energy bands and simulated several models in XSPEC to
 reproduce these numbers. In the observed spectrum we found that 12 source
 photons were at energies 
 E$<$ 2.0 keV, 24 source photons at E$>$2.0 keV (Table\,\ref{xrt_stat}), 
 16 photons between 2 and 5 keV
 and 8 photons between 5 and 10 keV. The spectrum that can reproduce these
 numbers the best is an absorbed two power law model $wabs \times (zwabs \times
 powl + zwabs \times powl)$. The absorption column density at z=0 was set to
 the Galactic value. The column densities of the redshifted absorbers are $5
 \times 10^{21}$ and $2 \times 10^{23}$ cm$^{-2}$ for the first and second
 power law component, respectively and X-ray spectral slopes \ax=8.7 and \ax=3.0
 for these components. The first power law spectral slope is that found during
 the RASS \citep{grupe95}. Even though this model can reproduce the observed
 photon distribution, a more physical model is a partial covering absorber with
 a single power law. We found that a partial covering absorber with a column
 density $N_{\rm H} = 1\times 10^{23}$ cm$^{-2}$ and a high covering fraction of
 $f_{\rm pc}$=0.95 and an X-ray spectral slope \ax=1.5 can also represent the
 observed photon distribution quite well. 
 Figure\,\ref{wpvs007_xspec} shows a simulated 85 ks spectrum (red triangles) 
 using the partial
 covering absorber model as described above.

\subsection{UVOT data}

\begin{figure}
\epsscale{0.6}
\plotone{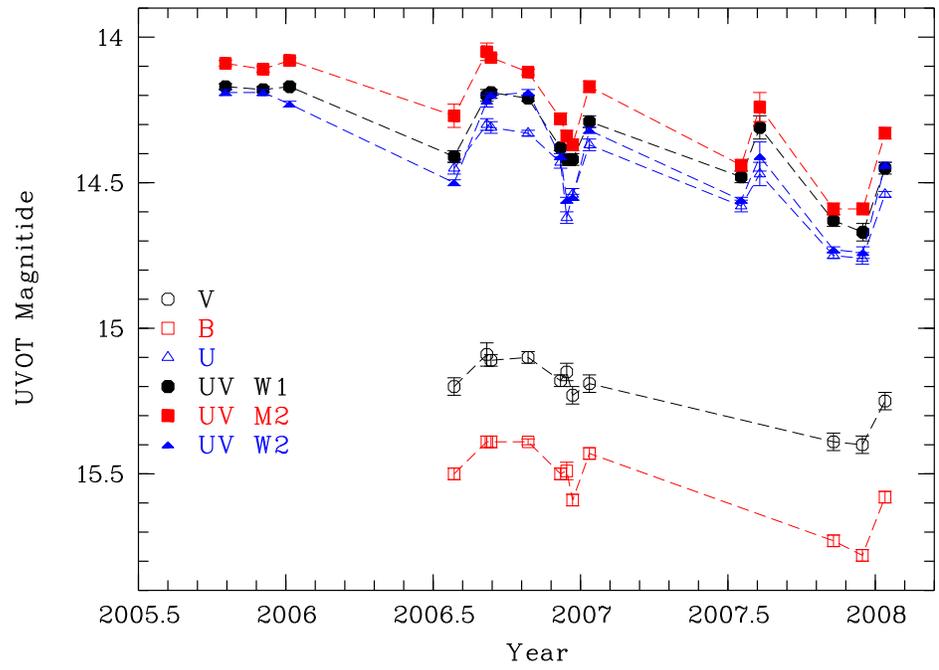}
\caption{\label{wpvs007_uvot_lc} UVOT light curves of WPVS 007 starting in 2005
October.
}
\end{figure}

Figure\,\ref{wpvs007_uvot_lc} displays the UVOT light curves in all 6 filters.
Note that during some of the observations only the UV filters have been used.
These light curves are similar to those shown in \citet{grupe07}, 
but all UVOT
data have been re-analyzed due to the most recent software and calibration
changes \citep{poole07}. WPVS 007 continues to be variable by about 0.3 mag in
the UV on timescales of months. The UV light curves also suggest that over the
period of about two years the AGN has become fainter by up to 0.5 mag (see
also Table\,\ref{uvot_log}) in the UV filters.

Because we can finally see hard X-ray photons in WPVS 007, it is possible to
construct the spectral energy distribution (SED) and derive the optical to X-ray
spectral slope \aox. The photon distribution of WPVS
007 presented here is dominated by the 2007 September observations when it
appeared to be brighter compared to the previous observations. Taking this
into account we measure an optical to X-ray spectral slope \aox=2.5. This is
the \aox\ of the absorbed/reddened spectrum. Note that
\citet{leighly08} found an unusual reddening curve of WPVS 007 based on the HST
data in comparison with HST spectra of Mrk 493 and Mrk 335.
In order to estimate the unabsorbed \aox,
we assumed an optical/UV spectrum of an (almost) unabsorbed NLS1. Here we picked
the \swift\ observation of MS 0117-28 (Grupe et al. 2008 in prep). For the
X-ray spectrum we used the unabsorbed power law spectrum found from the
partial covering absorber model fit with \ax=1.5. Based on these assumptions
we found that the unabsorbed optical to X-ray spectral slope is in the order
of \aox=1.9. The UV spectrum of MS 0117-28 can also be used to estimate the
approximate intrinsic reddening in WPVS 007. We measured a reddening of about
1 magnitude in the UVW2 filter which translates to an intrinsic reddening of
$E_{\rm B-V}$=0.1 mag.  
 This SED results in a
Big-Blue-Bump luminosity of log $L_{\rm BBB}$=37.7 [W] which is exactly the
Eddington luminosity of a $4 \times 10^6$ \msun black hole\footnote{The black
hole mass was determined using the relation given in \citep{kaspi00}.}, 
or in other words, the
Eddington ratio is $L/L_{\rm Edd}\approx1$, which is a typical value for NLS1s
\citep[e.g.][]{grupe04}.

\section{\label{discuss} Discussion}

We presented new \swift\ observations on the X-ray transient NLS1 \wpvs. 
We were able to detect the AGN in X-rays at a level of 5$\times 10^{-4}$ 
XRT counts s$^{-1}$ in the 0.3-10.0 keV band which is equivalent to an X-ray
flux of 2$\times 10^{-17}$ W m$^{-2}$. 
The most interesting aspect of this detection is that the majority of source
photons are detected at energies $E>$2 keV which is extremely surprising
considering that WPVS 007 used to be the AGN with the softest X-ray spectrum in
the sky, having no photon above 0.5 keV detected during the RASS.
 
Compared with the ACIS-S 
count rate measured during the 2002 \chandra\ observation 
\citep[$1\times 10^{-3}$ ACIS-S counts s$^{-1}$;][]{vaughan04} 
this 
indicates that WPVS 007 has become brighter by a factor of a few. 
This variability, however, only occurs in the hard band. The 0.2-2.0 keV flux
between the \chandra\ 2002 and \swift\ 2007 observations remains constant
within the errors.
The
non-variability of the optical spectra seen by \citet{winkler92} and
\citet{grupe95} 
suggests that the intrinsic continuum spectrum may not 
have changed. This non variability in the optical spectrum is somewhat
similar to Mrk 335 where the optical spectra taken in 1999 during the X-ray high
state and 2007 September during the low-state are identical \citep{grupe08}.
The non variability of the optical spectrum together
with the strong variability seen in X-rays and the UV may suggest that the
inclination angle in which we observe the nuclear region is 
 rather high assuming that the variability is a consequence of
   variable absorption.  
 This is expected for BAL QSOs according to the AGN geometry model
 presented by \citet{elvis00}. However, this is not exclusive. Recently there
 have been reports by e.g. \citet{berrington07} and \citet{ghosh08} of BAL QSOs
 with polar outflows.

Although the best-fit model is the partial covering absorber model, in principle
a reflection model \citep[e.g.][]{fabian89,fabian04} results in a similar shape
of the X-ray spectrum. While it is impossible with the current data set to
distinguish betwen the partial covering absorber and refelction models, the
presence of a strong absorber in the UV spectrum \citep{leighly08} makes the
partial covering absorber model more plausible as the explanation for the
observed \swift\ X-ray spectrum.

One
question is if the X-ray spectrum has changed intrinsically, or whether the absorber has
become more transparent sometime 
 between 2002 and 2007. The strong UV absorption in the FUSE observations also
 suggests that WPVS 007 was highly absorbed in 2003 November. 
 There was only one photon found
above 2 keV in the \chandra\ 2002 observation, but 24 in the combined \swift\
data.
In order to answer the question we simulated a 10 ks ACIS-S spectrum
using the spectral parameters of a partial covering absorber fit derived from
the \swift\ data. Using these parameters we expect to see 13 photons with
energies above 2 keV in the \chandra\ data. However, only one single photon
with $E>$2keV was found. This is a statistically significant lower value than
the expected number of photons. We performed another simulation assuming a 
column density of the partial covering absorber of $N_{\rm H,pc} = 1\times
10^{24}$ cm$^{-2}$ and leaving all other spectral parameters at the same values.
This simulation showed that the hard X-ray photons could be suppressed
significantly down to one photon above 2 keV in a 10 ks Chandra ACIS-S observations. 
Even though we are dealing with low number photon statistics, 
the hardness ratios found from the \swift\ data prior 2007
January observations the Chandra 2002 data both indicate a rather soft X-ray
spectrum, while the 2007 September data suggest a rather hard spectrum. 
Because the flux during the 2007 September \swift\ observation is a factor of
about 4 higher than during the 2002 \chandra observation, it suggests that the
column density of the
absorber of the high energy spectrum had decreased significantly
sometime in the time between 2007 January and September.  This
result suggests that the absorber has started to disappear again. 
Note, that a change in the absorber column density of a fully-covered absorber
has exactly the opposite effect on the hardness ratio than what a partial
covered absorber does: while in increase of the absorption column density of a
fully-covered absorber causes the hardness ratio to become harder, because
primarily soft X-ray photons are absorbed, an increase of the absorption column
density of a partial covering absorber causes the source to appear softer,
because the soft X-ray photons are dominated by the 
unabsorbed fraction of the X-ray continuum
 while the hard X-ray photons become suppressed.  
Considering that we observed the flux to decrease significantly
between the RASS observation in 1990 and the pointed ROSAT observation
in 1993, we may suggest that we will be able to see WPVS 007 as a
bright X-ray NLS1 again within the next few years. 

The optical to X-ray spectral slope \aox=2.5 clearly classifies WPVS 007 as an
X-ray weak AGN following the definition by \citet{brandt00} who define 'X-ray
weak' AGN by \aox$\ge$2.0. The reddening-corrected
 \aox=1.9 is at the borderline. Note
that this value has large uncertainties. Nevertheless, this result is consistent
with the findings by \citet{leighly04} that NLS1s with outflows seen in the
blueshifted emission lines have steeper \aox\ than NLS1s without outflows. 

One question remains: why was the X-ray spectral slope during the RASS
observation so steep? An energy index of \ax=8.7 as measured from a
single absorbed power law model to the RASS spectrum \citep{grupe95} can not
be explained by standard accretion disk models. We can also exclude a tidal
disruption event by a star orbiting closely to the central black hole. First,
the optical to X-ray flux ratio during the RASS was log 
$f_{\rm X}/f_{\rm o}$=0.22\footnote{The typical optical to X-ray spectral slope
\aox\ cannot be applied to these data because no X-ray photons with $E>$0.5 keV
were detected in the RASS data making any estimate of \aox very uncertain.} 
which is typical for a Seyfert
1 galaxy \citep{beu99,maccacaro88},
and second we do not detect any significant changes in the optical spectrum.
This is different than what has been found in the X-ray transient Seyfert 2
galaxy IC 3599 \citep{grupe95a, brandt95} or as recently reported for 
SDSS J095209.56+214313.3  by \citet{komossa08}
where strong changes in the optical
emission lines were found as a result of an X-ray outburst caused by a
dramatic accretion event. Note also that the X-ray to optical flux ratio of 
IC 3599 during the RASS was log $f_{\rm X}/f_{\rm o}$=1.45. 
One possible explanation for the absence of any photon above 0.5 keV during the
RASS observation is the presence of an ionized, 'warm', absorber in the line of
sight. This type of absorber is transparent at soft energies but absorbs at
intermediate energies, and therefore provides an efficient
mechanism to produce very steep soft X-ray spectra
(e.g., Komossa \& Meerschweinchen 2000, Done \& Nayakshin 2007). 
Using {\it absori} within XSPEC we simulated a 300s
spectrum of WPVS 007 during the RASS with an absorber column density $N_{\rm
H}=5\times 10^{22}$ cm$^{-2}$ of the warm absorber with an ionization parameter
$\xi$=1000. Fitting this simulated spectrum with an absorbed  single power law model
results in \ax=6.2$^{+2.0}_{-1.2}$, which is similar to what has been found
during the RASS \citep{grupe95}.

Our detection of WPVS 007 with \swift\ and the extraction of a photon
distribution from the 2005 to 2007 data opens a new window in our understanding
of WPVS 007 and X-ray weak AGN in general.
While in the UV we were able to follow the increase of the
absorber column density \citep{leighly05, leighly08} we are now able to follow
the disappearance of the absorber in X-rays. This will need future follow-up
observations by \xmm\ or \chandra\ and \swift. While \xmm\ or \chandra\ 
will be able to obtain a detailed
low-state spectrum, \swift\ will continue monitoring \wpvs\ in order to
determine when it will become X-ray bright again. In addition, optical
polarimetry or spectropolarimetry is needed to 
measure how much of the optical emission
is seen directly and how much is scattered (polarized) emission. Spectropolarimetry
will put strong constraints on the geometry of the AGN and will confirm (or not) the
assumption that WPVS 007 is viewed at a rather high inclination angle. 
We also have an approved HST COS observation along with a Chandra observation
in order to study the developments in the UV absorption lines and in the X-ray
continuum.  

\acknowledgments

First we want to thank Neil Gehrels for approving our ToO requests and
the \swift\ team for performing the ToO observations of
WPVS 007 and scheduling the AGN on a regular basis.
Many thanks also to Kim Page and the anonymous referee
for carefully reading the manuscript and helpful suggestions to improve the
paper.
This research has made use of the NASA/IPAC Extragalactic
Database (NED) which is operated by the Jet Propulsion Laboratory,
Caltech, under contract with the National Aeronautics and Space
Administration.
\swift\ at PSU is supported by NASA contract NAS5-00136.
This research was also supported by NASA contract NNX07AH67G (D.G.).

\clearpage



\clearpage

\begin{deluxetable}{lccr}
\tablecaption{\swift~XRT Observation log of WPVS 007
\label{xrt_log}}
\tablewidth{0pt}
\tablehead{
\colhead{Segment} & \colhead{T-start\tablenotemark{1}} & 
\colhead{T-stop\tablenotemark{1}} &
\colhead{$\rm T_{exp}$\tablenotemark{2}} 
} 
\startdata
014 & 2007-07-15 01:35 & 2007-07-16 17:42 & 2372 \\
015 & 2007-07-18 19:21 & 2007-07-18 19:46 & 1473 \\
016 & 2007-08-10 14:13 & 2007-08-10 14:15 &  145 \\
018 & 2007-09-06 23:03 & 2007-09-06 23:10 &  500 \\
019 & 2007-09-08 00:40 & 2007-09-09 23:28 & 15830 \\
020 & 2007-09-10 00:51 & 2007-09-10 23:34 & 7122 \\
021 & 2007-09-11 01:01 & 2007-09-11 23:40 & 6118 \\
022 & 2007-09-12 00:52 & 2007-09-12 23:32 & 5822 \\
023 & 2007-09-13 01:15 & 2007-09-13 23:52 & 7197 \\
024 & 2007-09-14 01:17 & 2007-09-14 24:00 & 7942 \\
026 & 2007-10-07 00:44 & 2007-10-07 13:39 & 2155 \\
027 & 2007-10-08 00:50 & 2007-10-08 13:45 & 2275 \\
028 & 2007-11-09 00:23 & 2007-11-09 02:29 & 2270 \\
029 & 2007-12-15 10:09 & 2007-12-15 23:18 & 1790 \\
030 & 2008-01-12 17:56 & 2008-01-12 22:55 & 1528  
\enddata

\tablenotetext{1}{Start and End times are given in UT}
\tablenotetext{2}{Observing time given in s}
\end{deluxetable}

\begin{deluxetable}{lrcrcrcrcrcrc}
\tabletypesize{\tiny}
\tablecaption{\swift~UVOT Observation of WPVS 007
\label{uvot_log}}
\tablewidth{0pt}
\tablehead{
& \multicolumn{2}{c}{V} 
& \multicolumn{2}{c}{B} 
& \multicolumn{2}{c}{U}  
& \multicolumn{2}{c}{UV W1} 
& \multicolumn{2}{c}{UV M2} 
& \multicolumn{2}{c}{UV W2}  \\
\colhead{\rb{Segment}} & 
\colhead{$\rm T_{exp}$\tablenotemark{1}} & \colhead{$\rm Mag_{corr}$\tablenotemark{2}} &
\colhead{$\rm T_{exp}$\tablenotemark{1}} & \colhead{$\rm Mag_{corr}$\tablenotemark{2}} &
\colhead{$\rm T_{exp}$\tablenotemark{1}} & \colhead{$\rm Mag_{corr}$\tablenotemark{2}} &
\colhead{$\rm T_{exp}$\tablenotemark{1}} & \colhead{$\rm Mag_{corr}$\tablenotemark{2}} &
\colhead{$\rm T_{exp}$\tablenotemark{1}} & \colhead{$\rm Mag_{corr}$\tablenotemark{2}} &
\colhead{$\rm T_{exp}$\tablenotemark{1}} & \colhead{$\rm Mag_{corr}$\tablenotemark{2}} 
} 
\startdata
001 & \nodata & \nodata & \nodata & \nodata & \nodata & \nodata & 646    & 14.17\plm0.01 & 675  & 14.09\plm0.01 & 686  & 14.19\plm0.01 \\
002 & \nodata & \nodata & \nodata & \nodata & \nodata & \nodata & 550    & 14.18\plm0.01 & 588  & 14.11\plm0.01 & 588  & 14.19\plm0.01 \\
003 & \nodata & \nodata & \nodata & \nodata & \nodata & \nodata & 1056   & 14.17\plm0.01 & 1171 & 14.08\plm0.01 & 1171 &  14.23\plm0.01 \\
004 & 155 & 15.20\plm0.03 & 159 & 15.50\plm0.02 & 159 & 14.45\plm0.02 & 319  & 14.41\plm0.02 &  118 & 14.27\plm0.03 & 615  & 14.50\plm0.01 \\
005 &  55 & 15.09\plm0.04 & 170 & 15.39\plm0.02 & 170 & 14.30\plm0.02 & 340  & 14.20\plm0.02 &  144 & 14.07\plm0.03 & 392  & 14.22\plm0.02 \\
006 & 194 & 15.11\plm0.03 & 194 & 15.39\plm0.02 & 194 & 14.31\plm0.02 & 387  & 14.19\plm0.01 &  536 & 14.07\plm0.01 & 777  & 14.20\plm0.01 \\
007 & 344 & 15.10\plm0.02 & 336 & 15.39\plm0.01 & 335 & 14.33\plm0.01 & 686  & 14.21\plm0.01 &  767 & 14.12\plm0.01 & 767  & 14.19\plm0.01 \\
009 & 245 & 15.18\plm0.02 & 245 & 15.50\plm0.02 & 245 & 14.43\plm0.02 & 486  & 14.36\plm0.01 &  621 & 14.28\plm0.02 & 978  & 14.41\plm0.01 \\
010 & 134 & 15.15\plm0.03 & 134 & 15.49\plm0.03 & 134 & 14.62\plm0.02 & 267  & 14.42\plm0.01 &  376 & 14.34\plm0.02 & 534  & 14.56\plm0.01 \\
011 & 187 & 15.23\plm0.03 & 187 & 15.59\plm0.02 & 187 & 14.54\plm0.02 & 374  & 14.42\plm0.02 &  513 & 14.37\plm0.02 & 750  & 14.55\plm0.01 \\
012+013 & 155 & 15.19\plm0.03 & 155 & 15.43\plm0.02 & 155 & 14.37\plm0.02 & 312 & 14.29\plm002 & 373 & 14.17\plm0.02 & 625 & 14.32\plm0.01\\
014 & \nodata & \nodata & \nodata & \nodata &      19 & 14.65\plm0.05 &  37  & 14.46\plm0.05 &  111 & 14.47\plm0.04 &  50  & 14.65\plm0.05 \\
015 & \nodata & \nodata & \nodata & \nodata &     145 & 14.58\plm0.02 & 290  & 14.48\plm0.02 &  436 & 14.44\plm0.02 & 560  & 14.56\plm0.01 \\
016 & \nodata & \nodata & \nodata & \nodata &      16 & 14.47\plm0.04 &  31  & 14.31\plm0.04 &   46 & 14.24\plm0.05 &  35  & 14.41\plm0.05 \\
028 & 189 & 15.39\plm0.03 & 189 & 15.73\plm0.02 & 187 & 14.75\plm0.01 & 374  & 14.63\plm0.02 &  510 & 14.59\plm0.02 & 745  & 14.73\plm0.01 \\        
029 & 152 & 15.40\plm0.03 & 153 & 15.78\plm0.02 & 153 & 14.76\plm0.02 & 124  & 14.67\plm0.03 &  375 & 14.59\plm0.02 & 248  & 14.74\plm0.02 \\
030 & 144 & 15.25\plm0.03 & 156 & 15.58\plm0.02 & 156 & 14.54\plm0.01 & 310  & 14.45\plm0.02 &  260 & 14.33\plm0.02 & 623  & 14.44\plm0.01  
\enddata
\tablenotetext{1}{Observing time given in s}
\tablenotetext{2}{Magnitude corrected for reddening with $E_{\rm B-V}$=0.012 given by
\citet{sfd98}. The errors given in this table are statistical errors}
\end{deluxetable}

\begin{deluxetable}{lcccccccc}
\tabletypesize{\scriptsize}
\tablecaption{Count statistics of the WPVS 007 \swift\ observations
\label{xrt_stat}}
\tablewidth{0pt}
\tablehead{
& & \multicolumn{2}{c}{source+backgr} &
\multicolumn{2}{c}{background\tablenotemark{2}} \\
\colhead{\rb{Observations}} & \colhead{\rb{$T_{\rm exp}$\tablenotemark{1}}} & 
\colhead{S\tablenotemark{3}} & \colhead{H\tablenotemark{3}} &
\colhead{S\tablenotemark{3}} & \colhead{H\tablenotemark{3}} &
\colhead{counts} &
\colhead{CR\tablenotemark{4}} &
\colhead{HR\tablenotemark{3}} 
} 
\startdata
2005 October - 2007 January & 23105 & 7 & 4 & 30 & 33 & 7.2$^{+2.7}_{-3.8}$ &
3.1$^{+1.2}_{-1.6}$ & $-0.49^{+0.23}_{-0.50}$ \\
2007 September & 50531 & 8 & 24 & 46 & 50 & 25.9$^{+6.1}_{-6.3}$ & 5.1\plm1.2 &
+0.63\plm0.20 \\
2005 October - 2008 January & 85508 & 18 & 30 & 92 & 105 & 35.7$^{+6.4}_{-6.7}$ & 4.2\plm0.8 &
+0.32\plm0.19 
\enddata

\tablenotetext{1}{Exposure times are given in s}
\tablenotetext{2}{The background area is 16 times larger than the source area}
\tablenotetext{3}{The hardness ratio is defined as HR=(H-S)/(H+S) with S and H are
background subtracted counts in the 0.3-2.0 and 2.0-10.0 keV bands, respectively. The
hardness ratio was calculated following the description in \citet{park06}. }
\tablenotetext{4}{Count rate is given in units of $10^{-4}$ counts s$^{-1}$.}

\end{deluxetable}

\end{document}